%% file: b_print.tex
\documentclass[aps,prl,showpacs,superscriptaddress,preprintnumbers,amsmath,amssymb]{revtex4}

\usepackage{graphicx} 
\usepackage{subfigure}

\newcommand{\cc}{\ensuremath{(c\bar{c})_{\mathrm{res}}}}
\newcommand{\ncc}{\ensuremath{(c\bar{c})_{\mathrm{non-res}}}}
\newcommand{\RM}{\ensuremath{M_{\mathrm{recoil}}}}
\newcommand{\aprod}{\ensuremath{\alpha_{\text{prod}}}}
\newcommand{\ahel}{\ensuremath{\alpha_{\text{hel}}}}
\newcommand{\tprod}{\ensuremath{\theta_{\text{prod}}}}
\newcommand{\thel}{\ensuremath{\theta_{\text{hel}}}}

\begin{document}

\includegraphics[width=3cm] {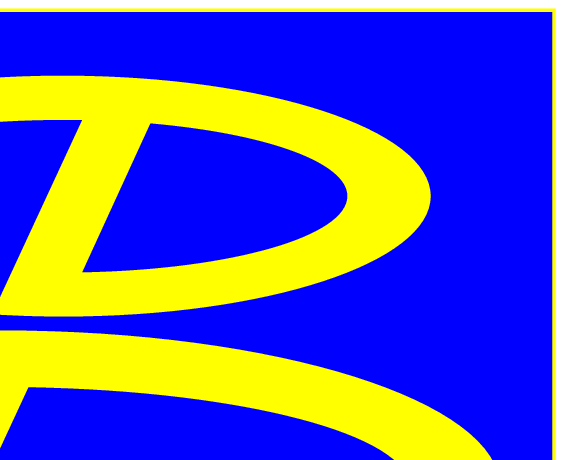}
\noindent
\title{ Study of double charmonium production in $e^{+}e^{-}$
annihilation at $\sqrt{s} \approx 10.6~$GeV}

\input{author.tex}

\begin{abstract}
\noindent
We present a new analysis of double charmonium production in $e^+ e^-$
annihilation.  The observation of the processes $e^+ e^- \to J/\psi \,
\eta_c$, $J/\psi\,\chi_{c0}$, and $J/\psi\,\eta_c(2S)$ is confirmed
using a dataset more than three times larger than that of Belle's
previous report, and no evidence for the process $e^+ e^- \to J/\psi
\, J/\psi$ is found. We perform an angular analysis for
$J/\psi\,\eta_c$ production and set an upper limit on the production
of $J/\psi\,J/\psi$.  Processes of the type $e^+ e^- \to \psi(2S) \cc$
have been observed for the first time; their rates are found to be
comparable to those of $e^+ e^- \to J/\psi \cc$ processes.

\end{abstract}


\maketitle
\setcounter{footnote}{0}

\noindent
The large rate for processes of the type $e^+ e^-\to J/\psi\,\eta_c$
and $J/\psi\,\ncc$ reported by Belle~\cite{2cc} remains
unexplained. Following the publication of this result, the
cross-section for $e^+ e^-\to J/\psi\,\eta_c$ via $e^+ e^-$
annihilation into a single virtual photon was calculated using
non-relativistic QCD (NRQCD) to be $\sim 2
\,\mathrm{fb}$~\cite{bra_excl}, which is at least an order of
magnitude smaller than the measured value. Several hypotheses have
been suggested in order to resolve this discrepancy. In particular,
the authors of Ref.~\cite{bra_psi} have proposed that processes
proceeding via two virtual photons may be important. Other
authors~\cite{gold} suggest that since the dominant mechanism for
charmonium production in $e^+e^-$ annihilation is expected to be the
color-singlet process $e^+e^- \to c\bar{c} g g$, the final states
observed by Belle contain a charmonium state and a $M \sim
3\,\mathrm{GeV}/c^2$ glueball. Such glueball states are predicted by
lattice QCD and can have masses around $3\,\mathrm{GeV}/c^2$. Possible
glueball contributions to the $\chi_{c0}$ signal are also discussed in
Ref.~\cite{hagi}.

The previous Belle analysis was performed with a data sample of
$45\,$fb$^{-1}$. The process $e^+ e^-\to J/\psi\,\eta_c$ was inferred
from the $\eta_c$ peak in the mass spectrum of the system recoiling
against the reconstructed $J/\psi$ in inclusive $e^+ e^- \to
J/\psi\,X$ events. In this paper we report an extended analysis of the
$e^+ e^- \to J/\psi \, \cc$ process to check the above hypotheses and
provide extra information that might be useful to resolve the puzzle.
This study is performed using a data sample of $140
\,\mathrm{fb}^{-1}$ collected at the $\Upsilon(4S)$ resonance and
$15\,\mathrm{fb}^{-1}$ at an energy $60\,{\mathrm{MeV}}$ below the
$\Upsilon(4S)$. The data were collected with the Belle
detector~\cite{Belle} at the KEKB asymmetric energy $e^+ e^-$ storage
rings~\cite{KEKB}.

The analysis procedure is described in detail in Ref.~\cite{2cc}. For
$J/\psi$ reconstruction we combine oppositely charged tracks that are
both positively identified either as muons or electrons.  For $J/\psi
\to e^{+}e^{-}$, the invariant mass calculation includes the
four-momentum of photons detected within $50\,\mathrm{mrad}$ of the
$e^{\pm}$ directions, as a partial correction for final state
radiation and bremsstrahlung energy loss. The $J/\psi \to
\ell^{+}\ell^{-}$ signal region is defined by a mass window
$\left|M_{\ell^{+}\ell^{-}} - M_{J/\psi}\right| < 30 \,
\mathrm{MeV}/c^2$ ($\approx 2.5\, \sigma_M$).  QED processes are
significantly suppressed by the requirement that the total charged
multiplicity ($N_{\rm ch}$) in the event be $N_{\rm ch}>4$.  The
contribution from $J/\psi$ mesons in $B\overline{B}$ events is removed
by requiring the center-of-mass (CM) momentum $p^{*}_{J/\psi}$ to be
greater than $2.0\,\mathrm{GeV}/c$.  A mass-constrained fit is then
performed to improve the $p^{*}_{J/\psi}$ resolution and the recoil
mass $\RM = \sqrt{(E_{\rm CM}-E_{J/\psi}^*)^2-p_{J/\psi}^{*~2}}$ is
calculated, where $E_{J/\psi}^*$ is the $J/\psi$ CM energy after the
mass constraint.  $\psi(2S)$ is reconstructed via its decay to
$J/\psi\,\pi^+ \pi^-$ and the $\psi(2S)$ signal window is defined as
$|M_{J/\psi\, \pi^+ \pi^-} -M(\psi(2S))| <10\,\mathrm{MeV}/c^2$
($\approx 3\, \sigma_M$).

\begin{figure}[htb]
\hspace*{-0.025\textwidth}
\includegraphics[width=0.75\textwidth] {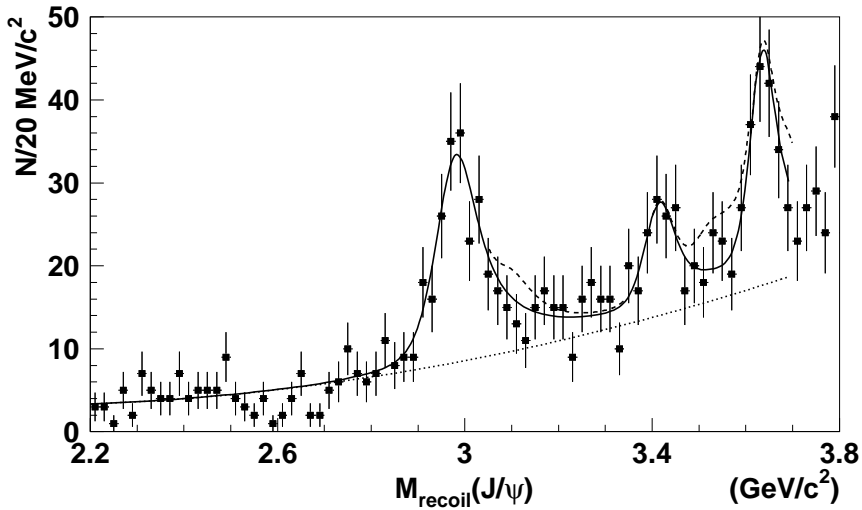}
\caption{The mass of the system recoiling against the
reconstructed $J/\psi$ in inclusive $e^+e^- \to J/\psi\,X$ events. The
curves are described in the text.}
\label{cc7}
\end{figure}

The $\RM(J/\psi)$ spectrum for the data is presented in
Fig.~\ref{cc7}: clear peaks around the nominal $\eta_c$
and $\chi_{c0}$ masses are evident; another significant peak around
$\sim 3.63\,\mathrm{GeV}/c^2$ is identified as the $\eta_c(2S)$.  The
authors of Ref.~\cite{bra_psi} estimated that the two-photon-mediated
process $e^+e^-\to J/\psi\, J/\psi$ has a significant cross-section
and suggested that the observed $e^+ e^- \to J/\psi\, \eta_c$ signal
in~\cite{2cc} might also include double $J/\psi$ events, thereby
producing an inflated cross-section measurement. Since $e^+e^-$
annihilation to $J/\psi\, J/\psi$ via a single virtual photon is
forbidden by charge conjugation symmetry, it was ignored in our
previous analysis. To allow for a possible contribution from the
exchange of two virtual photons, we fit the spectrum in Fig.~\ref{cc7}
including all of the known narrow charmonium states. In this fit, the
mass positions for the $\eta_c$, $\chi_{c0}$ and $\eta_c(2S)$ are free
parameters; those for the $J/\psi$, $\chi_{c1}$, $\chi_{c2}$ and
$\psi(2S)$ are fixed at their nominal values.  The expected
line-shapes for these peaks are determined from a Monte Carlo (MC)
simulation as described in our previous paper~\cite{2cc}, the
background is parameterized by a second order polynomial function, and
only the region below the open charm threshold ($M_{\rm recoil} < 3.7
\, \mathrm{GeV} /c^2$) is included in the fit.  The fit results are
listed in Table~\ref{cc3t}. The yields for $\eta_c$, $\chi_{c0}$, and
$\eta_c(2S)$ have statistical significances between 3.8 and 10.7.  The
significance of each signal is defined as $\sqrt{-2\ln (\mathcal{L}_0/
\mathcal{L}_{\text {max}})}$, where $\mathcal{L}_0$ and
$\mathcal{L}_{\text {max}}$ denote the likelihoods with the
corresponding signal yield fixed at zero and at the best-fit value,
respectively. The fit returns negative yields for the $J/\psi$ and
$\psi(2S)$; the $\chi_{c1}$ and $\chi_{c2}$ yields are found to be
consistent with zero. A fit with all these contributions fixed at zero
is shown as a solid line in Fig.~\ref{cc7}; the difference in the
$\eta_c$, $\chi_{c0}$ and $\eta_c(2S)$ yields compared to the default
fit is small, and is included in the systematic errors.  The dashed
line in the figure corresponds to the case where the contributions of
the $J/\psi$, $\chi_{c1}$, $\chi_{c2}$ and $\psi(2S)$ are set at their
90$\%$ confidence level upper limit values. The dotted line is the
background function.
\begin{table}[htb]
\vbox {
\caption{Summary of the signal yields ($N$), charmonium masses ($M$),
significances, and cross-sections ($\sigma_{Born} \times
\mathcal{B}_{>2}(\cc)$) for $e^+ e^- \to J/\psi \, \cc$;
$\mathcal{B}_{>2}$ denotes the branching fraction for final states
with more than two charged tracks.
}
\label{cc3t}
\begin{center}
\begin{ruledtabular}
\begin{tabular}{lrccc}
\cc & \multicolumn{1}{c}{$N$} & $M\,[\mathrm{GeV}/c^2]$& Signif.\ &
$\sigma_{Born}\! \! \times \!\mathcal{B}_{>2}\; [\mathrm{fb}]$ \\ \hline

$  \eta_c$ & $ 235 \pm 26 $ & $ 2.972 \pm 0.007 $ & 
$  10.7  $ & $25.6\pm 2.8\pm 3.4$ \\

$  J/\psi$ & $-14 \pm 20$ & fixed & --- & \multicolumn{1}{r}{$<9.1$ at 90\% CL} \\

$ \chi_{c0}$ & $89 \pm 24 $ & $3.407 \pm 0.011 $ & $3.8$ & $\phantom{2}6.4\pm
1.7\pm 1.0$ \\

$ \chi_{c1}\!+\!\chi_{c2} $ & $10 \pm 27$ & fixed & --- & 
\multicolumn{1}{r}{$<5.3$ at 90\% CL} \\

$ \eta_c(2S)$ & $164 \pm 30$ & $3.630 \pm 0.008 $ & $6.0$ & $16.5\pm
3.0 \pm 2.4$ \\

$ \psi(2S)$ & $-26 \pm 29$ & fixed & --- &  \multicolumn{1}{r}{$<13.3$ at 90\% CL}
\end{tabular}
\end{ruledtabular}
\end{center}
}
\end{table}

\begin{figure}[htb]
\hspace*{-0.025\textwidth}
\includegraphics[width=0.75\textwidth] {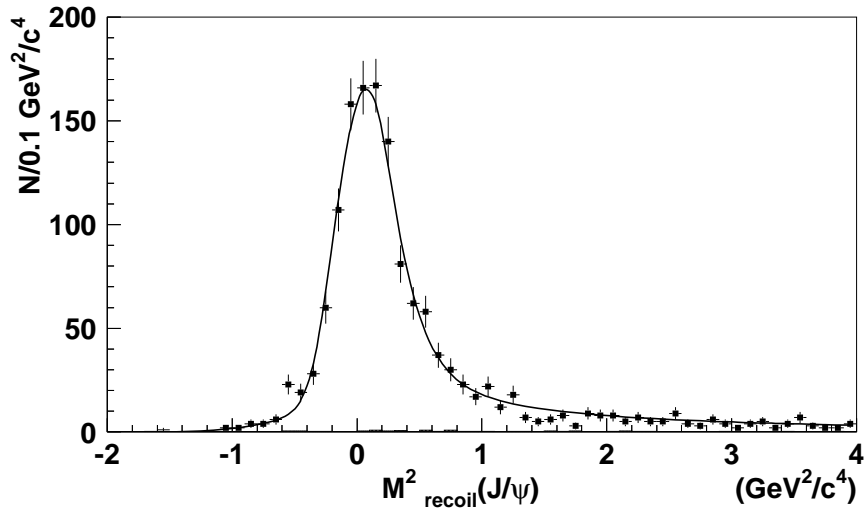}
\caption{The square of the mass of the system recoiling against the
	reconstructed $\psi(2S)\to J/\psi\, \pi^+ \pi^-$, in events
	with charged multiplicity equal to 4. Points with error bars
	show the data, the solid line shows the result of the fit
	described in the text.  The hatched histogram (scarcely
	visible) shows the spectrum in the scaled $\psi(2S)$
	sidebands.}
\label{return}
\end{figure}

Given the arguments in Ref.~\cite{bra_psi}, it is important to check
for any momentum scale bias that may shift the recoil mass values and
confuse the interpretation of peaks in the \RM\ spectrum.  We use $e^+
e^-\to \psi(2S) \gamma$, $\psi(2S) \to J/\psi\,\pi^+\pi^-$ events to
calibrate and verify the recoil mass scale. Events with a
reconstructed $\psi(2S)$ and with no other charged tracks form a pure
$e^+ e^-\to \psi(2S) \gamma$ sample with less than $1\%$ background as
estimated using the $\psi(2S)$ sideband region.  We use the $\psi(2S)$
momentum to calculate the square of the mass of the recoiling system;
the resulting spectrum is shown in Fig.~\ref{return}.
The scaled $\psi(2S)$ sideband is also shown. We perform a fit to the
$\RM^2(\psi(2S))$ spectrum, using Monte Carlo simulation to determine
the expected signal shape; second order $QED$ corrections, which
produce a higher $\RM^2$ tail, are taken into account. The peak
position is left free in the fit and  the non-$\psi(2S)$ background is
ignored. The fit finds the shift in the data with respect to the MC
function to be consistent with zero ($\Delta \RM^2 = 0.010 \pm 0.009\,
\mathrm{GeV}^2/c^4$). From this result we conclude that the $J/\psi$
recoil mass is shifted by not more than $3\,\mathrm{MeV}/c^2$ in the
region $\RM(J/\psi)\sim 3\, \mathrm{GeV}/c^2$.

As an additional cross-check we fully reconstruct double charmonium
events. The $\eta_c$ is reconstructed as $K^0_S K^{\pm}
\pi^{\mp}~(K^0_S \to \pi^+ \pi^-)$ or $ 2(K^+ K^-)$ combinations
within a window of $\pm50\,\mathrm{MeV}/c^2$ around the nominal
$\eta_c$ mass. In events with $N_{\rm{ch}}=6$ we find 3 events with
$J/\psi\,\eta_c$ combinations in a $\pm 100\,\mathrm{MeV}$ window
around the CM energy ($\approx 3\, \sigma$).  No events are seen in
the $\eta_c$ sideband region ($100<M(K_S K \pi /2(KK)) -
M_{\eta_c}<350\,\mathrm{MeV}$); a fit to the mass distribution gives
an $\eta_c$ signal significance of $4.1\sigma$.  Based on the $\eta_c$
yield in the \RM$(J/\psi)$ distribution, we expect $2.6\pm 0.8$ fully
reconstructed events, consistent with the observed signal. Thus we
conclude that the peak in \RM$(J/\psi)$ is dominated by $\eta_c$
production. We also search for fully reconstructed double $J/\psi$
candidates in events with $N_{\rm{ch}}=4$.  No $J/\psi\,J/\psi$
candidates are found in a window of $\pm 100\,\mathrm{MeV}$ around the
CM energy.

Based on the calibration of the \RM$(J/\psi)$ scale, the result of the
fit to the $\RM(J/\psi)$ distribution and the full reconstruction
cross-check, we confirm our published observation of the process $e^+
e^- \to J/\psi\, \eta_c$ and rule out the suggestion of
Ref.~\cite{bra_psi} that a significant fraction of the inferred
$J/\psi \, \eta_c$ signal might be due to $J/\psi\,J/\psi$ events.

\begin{figure}[htb]
\hspace*{-0.025\textwidth}
\includegraphics[width=0.75\textwidth] {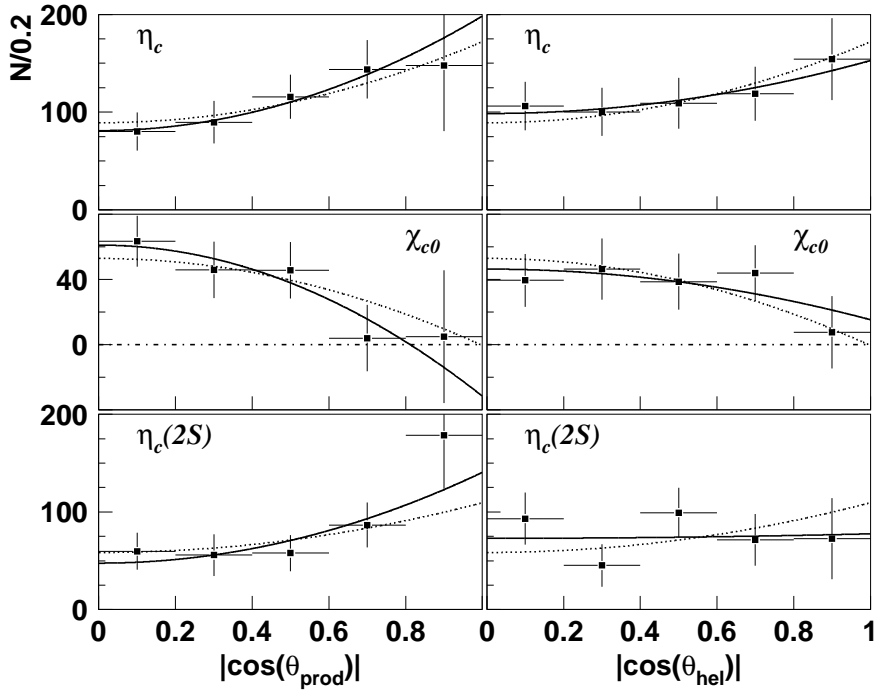}
\caption{Distributions of cosines of the production (left) and
$J/\psi$ helicity angles (right) for $e^+ e^- \to J/\psi\,\eta_c$ (top
row), $e^+ e^- \to J/\psi\,\chi_{c0}$ (middle row) $e^+ e^- \to
J/\psi\,\eta(2S)$ (bottom row). The solid lines are results of the
individual fits; the dotted lines are the simultaneous fit results.}
\label{angl}
\end{figure}

The reconstruction efficiencies for the $J/\psi\,\eta_c$,
$J/\psi\,\chi_{c0}$, and $J/\psi\,\eta(2S)$ final states strongly
depend on \tprod, the production angle of the $J/\psi$ in the CM frame
with respect to the beam axis, and the helicity angle \thel, defined
as the angle between the decay $\ell^+$ direction and the boost
direction of the CM frame in the $J/\psi$ rest frame.  We therefore
perform an angular analysis for these modes before computing
cross-sections. We fit the \RM$(J/\psi)$ distributions in bins of
$\left|\cos(\tprod)\right|$ and $\left|\cos(\thel)\right|$, and
correct the yield for the reconstruction efficiencies determined
bin-by-bin from the MC.  The results are plotted in Fig.~\ref{angl},
together with fits to functions $A(1+\alpha\cos^2{\theta})$ (solid
lines). We also perform simultaneous fits to the production and helicity
angle distributions for each of the $\cc$ states, assuming 
$J/\psi\, \cc$ production via a single virtual photon and
angular momentum conservation, thus setting $\aprod \equiv \ahel$. The
values of the parameter $\alpha$ from the separate fits to
$|\cos(\thel)|$ and $|\cos(\tprod)|$, and from the simultaneous fits,
are listed in Table~\ref{ang}.
\begin{table}[htb]
\vbox {
\caption{The $\alpha$ parameters obtained from fits to the production
and helicity angle distributions for $e^+ e^- \to J/\psi \, \cc$.}
\label{ang}
\begin{center}
\begin{ruledtabular}
\begin{tabular}{lccc}
 & \multicolumn{2}{c}{Separate fits} & Simultaneous fits \\
 \cc     & \, \aprod\ \, & \, \ahel\ \,  & \,
$\ahel \equiv \aprod$ \,\\
\hline
$\, \eta_c$     & $\, \phantom{-}1.4^{+1.1}_{-0.8}\, $  & 
$\, \phantom{-}0.5^{+0.7}_{-0.5}\, $ & $\, \phantom{-}0.93^{+0.57}_{-0.47}\, $\\
$\, \chi_{c0}$  & $\, -1.7\pm0.5\, $ & 
$\, -0.7^{+0.7}_{-0.5}\, $ & $\, -1.01^{+0.38}_{-0.33}\, $\\
$\, \eta_c(2S)\, $ & $\phantom{-}1.9^{+2.0}_{-1.2}\, $  & 
$\phantom{-}0.3^{+1.0}_{-0.7}$ & $\, \phantom{-}0.87^{+0.86}_{-0.63}\, $ \\
\end{tabular}
\end{ruledtabular}
\end{center}
}
\end{table}

The angular distributions for the $J/\psi\,\eta_c$ and $J/\psi\,\eta_c(2S)$ 
peaks are consistent with the expectations for production of these final
states via a single virtual photon, $\aprod = \ahel = +1$~\cite{bra_excl}.
There is no evidence for the sharp rise in cross-section for
large $|\cos(\tprod)|$ expected for $J/\psi\,J/\psi$ production via two
virtual photons~\cite{bra_psi}.
The prediction for a spin-0 glueball contribution
($e^+ e^- \to J/\psi\,\mathcal{G}_0$) to the $J/\psi\,\eta_c$ peak,
$\aprod = \ahel \simeq -0.87$~\cite{gold}, is also disfavored.

The process $e^+ e^- \to \gamma^\ast \to J/\psi\,\chi_{c0}$ can proceed
via both S- and D-wave amplitudes, and predictions for the resulting angular
distributions are therefore model dependent. Our results disfavor the 
NRQCD expectation $\aprod = \ahel \simeq 0.25$~\cite{bra_excl,hagi},
and are more consistent with $S$-wave production,
where $\aprod = \ahel = -1$.

To calculate the cross-sections for the processes $e^+ e^- \to
J/\psi\, \eta_c, \, J/\psi\, \chi_{c0}, \, J/\psi\, \eta(2S)$ we fix
the production and helicity angle distributions in the MC to
$1+\cos^2{\theta}$ for $J/\psi\, \eta_c(\eta_c(2S))$, and to
$1-\cos^2{\theta}$ for $J/\psi\, \chi_{c0}$. The statistical errors in
the $\alpha$ parameters for the angular distributions are translated
into uncertainties in the efficiency determination and included in the
systematic error. To set a conservative upper limit for $e^+ e^- \to
J/\psi\, J/\psi, \, J/\psi\, \chi_{c1(2)}, \, J/\psi\, \psi(2S)$, we
use assumptions for the production and helicity angle distributions
that correspond to the lowest detection efficiency. Note that for
$J/\psi\, J/\psi$ and $J/\psi\, \psi(2S)$ the assumed angular
distributions lead to lower efficiencies than those that follow from
the predictions of Ref.~\cite{bra_psi}.

To reduce the model dependence of our results due to the effect of
initial state radiation (ISR), whose form-factor dependence on $Q^2$
of the virtual photon is unknown, we calculate cross-sections in the
Born approximation. We first calculate the fraction of events in the
signal \RM$(J/\psi)$ distributions that are accompanied by an ISR
photon of an energy smaller than a cutoff $E_{\text{cutoff}}$ using
MC. We then correct the cross-sections calculated for that cutoff
value using a factor that yields the Born
cross-section~\cite{fad}. The final result ($\sigma_{Born}=0.70\cdot
\sigma_{full}$) is then independent of the choice of the cutoff energy
provided $E_{\text{cutoff}}$ satisfies $M_e \ll E_{\text{cutoff}} \ll
E_{CM}$. As in Ref.~\cite{2cc}, because of selection criteria we
present our result in terms of the product of the cross-section and
the branching fraction of the recoil charmonium state into more than 2
charged tracks: $\sigma \times \mathcal{B}_{>2}$, where
$\mathcal{B}_{>2}(\cc) \equiv \mathcal{B}(\cc \to \,
>2\,\text{charged})$. The cross-sections are given in
Table~\ref{cc3t}.

\begin{figure}[htb]
\hspace*{-0.025\textwidth}
\includegraphics[width=0.75\textwidth] {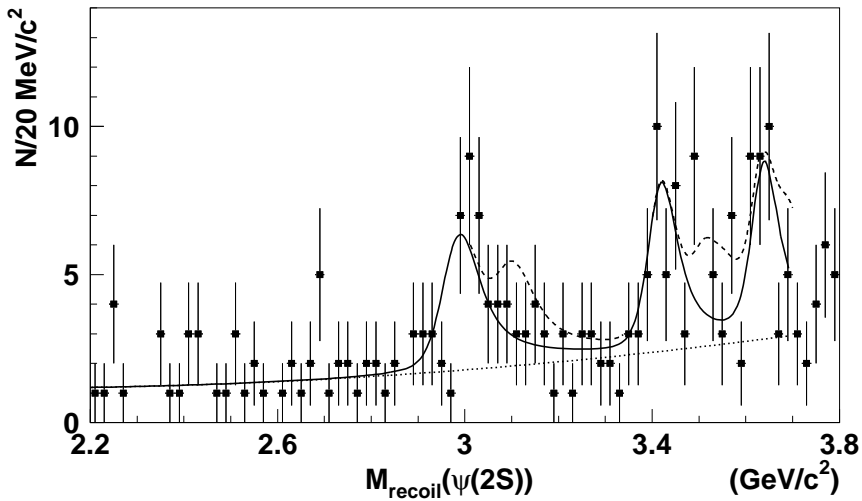}
\caption{The mass of the system recoiling against the
reconstructed $\psi(2S)$ in inclusive $e^+e^- \to \psi(2S) X$
events. The curves are described in the text.}
\label{cc7_2}
\end{figure}

We perform a similar study with reconstructed $\psi(2S) \to J/\psi\,
\pi^+ \pi^-$ decays to search for $e^+ e^- \to \psi(2S) \cc$ processes.
The recoil mass spectrum for the data is presented in
Fig.~\ref{cc7_2}: peaks corresponding to the $\eta_c$, $\chi_{c0}$,
and $\eta_c(2S)$ can be seen.
The fit to the $\RM(\psi(2S))$ distribution is identical to the
$\RM(J/\psi)$ fit, but due to the limited sample in this case, the
masses of the established charmonium states are fixed to their nominal
values; the $\eta_c(2S)$ mass is fixed to $3.630\,\mathrm{GeV}/c^2$ as
found from the $\RM(J/\psi)$ fit. The signal yields are listed in
Table~\ref{cc2_3t}.  Significances for the individual $\eta_c$,
$\chi_{c0}$, and $\eta_c(2S)$ peaks are in the range $3\sim4\sigma$;
the significance for $e^+e^- \to \psi(2S)\,\cc$, where $\cc$ is a sum
over $\eta_c$, $\chi_{c0}$, and $\eta_c(2S)$, is estimated to be
$5.3\,\sigma$. The significance is calculated as $\sqrt{-2\ln
(\mathcal{L}_0/ \mathcal{L}_{\text {max}})}$, where $\mathcal{L}_0$
and $\mathcal{L}_{\text {max}}$ denote the likelihoods of the fit with
all signal yields fixed at zero and at the best-fit value,
respectively.  In Fig.~\ref{cc7_2} the result of a fit with only
$\eta_c$, $\chi_{c0}$ and $\eta_c(2S)$ contributions included is shown
as a solid line; the dashed line shows the case where the $J/\psi$,
$\chi_{c1}$, $\chi_{c2}$, and $\psi(2S)$ contributions are set at
their $90\%$ confidence level upper limit values. The dotted line is
the background function.

To estimate the efficiency we assume the $\psi(2S)$ production and
helicity angle distributions to be the same as those for the
corresponding $J/\psi\, \cc$ final states. Finally, the calculated
products of the Born cross-section and the branching fraction of the
recoiling charmonium state into two or more charged tracks ($\sigma
\times \mathcal{B}_{>0}$, where $ \mathcal{B}_{>0}(\cc) \equiv
\mathcal{B}(\cc \to >\,0\,\text{charged})$) are presented in
Table~\ref{cc2_3t}.
\begin{table}[htb]
\vbox {
\caption{Summary of the signal yields ($N$), significances,
and cross-sections ($\sigma_{Born} \times
\mathcal{B}_{> 0}(\cc)$) for $e^+ e^- \to \psi(2S) \, \cc$;
$\mathcal{B}_{>0}$ denotes the branching fraction for final
states containing charged tracks.
}
\label{cc2_3t}
\begin{center}
\begin{ruledtabular}
\begin{tabular}{lccc}
\cc & $N$ & Signif. &
$\sigma_{Born}\times \mathcal{B}_{>0}\,[\mathrm{fb}]$ \\ \hline

$ \, \eta_c$ & $\, 36.7 \pm 10.4 \,$ & $ \, 4.2 \, $ & $16.3 \pm 4.6 \pm 3.9$\\

$\,  J/\psi$ & $6.9 \pm 8.9$ & --- & \multicolumn{1}{r}{$<16.9$ at 90\% CL}  \\

$\, \chi_{c0}$ & $35.4 \pm 10.7 $ & $3.5$ &  $12.5\pm 3.8 \pm 3.1$\\

$\, \chi_{c1}+\chi_{c2} \,$ & $6.6 \pm 8.0$  &  --- & \multicolumn{1}{r}{$<8.6$ at 90\% CL} \\

$\, \eta_c(2S)$ & $36.0 \pm 11.4$ &   $3.4$ & $16.0\pm 5.1 \pm 3.8$ \\

$\, \psi(2S)$ & $-8.3 \pm 8.5\phantom{-}$ & --- & \multicolumn{1}{r}{$<5.2$ at 90\% CL}  \\ 
\end{tabular}
\end{ruledtabular}
\end{center}
}
\end{table}

The systematic error is dominated by the fitting systematics of the
signal yields: $10\%$ for $J/\psi\, \cc$ and $14\%$ for
$\psi(2S)\cc$. To estimate this contribution we vary the
parameterizations of the signal (intrinsic widths of charmonium
states, form-factor dependence on $Q^2$) and background in the fit to
the \RM\ spectra. Another large contribution is due to the
reconstruction efficiency dependence on the angular distributions
($7\%$ for $J/\psi\, \cc$ and $15\%$ for $\psi(2S)\cc$). In the MC,
the angular parameters $\alpha$ are varied within the statistical
errors of our angular analysis for $e^+ e^- \to J/\psi \, \cc$ and in
the full range ($-1 \leq \aprod(\ahel) \leq 1$) for $e^+ e^- \to
\psi(2S) \, \cc$ to estimate the uncertainty in efficiencies. Other
contributions for $e^+ e^- \to J/\psi \cc$ ($\psi(2S) \cc$) come from
the multiplicity cut ($3\%(2\%)$), track reconstruction efficiency
($3\%(5\%)$) and lepton identification ($3\%(3\%)$).

In summary, using a larger data set we confirm our published
observation of $e^+ e^- \to J/\psi\, \eta_c, ~ J/\psi\, \chi_{c0}$ and
$J/\psi\, \eta_c(2S)$ and find no evidence for the process $e^+ e^-
\to J/\psi\, J/\psi$. We have calculated the cross-sections for $e^+
e^- \to J/\psi\, \eta_c$, $J/\psi\, \chi_{c0}$, and $J/\psi\, \eta_c(2S)$
with better statistical accuracy and reduced systematic errors and set
an upper limit for $\sigma(e^+ e^- \to J/\psi\, J/\psi) \times
\mathcal{B}(J/\psi \to \, >2 ~ \text{charged})$ of $9.1\,\mathrm{fb}$
at the $90\%$ CL. Although this limit is not
inconsistent with the prediction for the $J/\psi\, J/\psi$ rate given in
Ref.~\cite{bra_psi}, the suggestion that a large fraction of the inferred
$J/\psi \, \eta_c$ signal consists of $J/\psi\, J/\psi$ events is ruled out.
We have measured the production and helicity angle distributions for
$e^+ e^- \to J/\psi\,\eta_c$, $J/\psi\,\chi_{c0}$, and $J/\psi\,\eta_c(2S)$;
the distributions are consistent with expectations for these states, and
disfavor a spin-0 glueball contribution to the $\eta_c$ peak.
We observe $\psi(2S)\cc$
production for the first time, and find that the production rates for 
these final states are of the same magnitude as the corresponding rates
for $J/\psi\, \cc$.

We thank the KEKB group for the excellent
operation of the accelerator, the KEK Cryogenics
group for the efficient operation of the solenoid,
and the KEK computer group and the NII for valuable computing and
Super-SINET network support.  We acknowledge support from
MEXT and JSPS (Japan); ARC and DEST (Australia); NSFC (contract
No.~10175071, China); DST (India); the BK21 program of MOEHRD and the
CHEP SRC program of KOSEF (Korea); KBN (contract No.~2P03B 01324,
Poland); MIST (Russia); MESS (Slovenia); NSC and MOE (Taiwan); and DOE
(USA).

\end{document}

%% file: author.tex
\affiliation{Budker Institute of Nuclear Physics, Novosibirsk}
\affiliation{Chiba University, Chiba}
\affiliation{Chonnam National University, Kwangju}
\affiliation{University of Cincinnati, Cincinnati, Ohio 45221}
\affiliation{Gyeongsang National University, Chinju}
\affiliation{University of Hawaii, Honolulu, Hawaii 96822}
\affiliation{High Energy Accelerator Research Organization (KEK), Tsukuba}
\affiliation{Hiroshima Institute of Technology, Hiroshima}
\affiliation{Institute of High Energy Physics, Chinese Academy of Sciences, Beijing}
\affiliation{Institute of High Energy Physics, Vienna}
\affiliation{Institute for Theoretical and Experimental Physics, Moscow}
\affiliation{J. Stefan Institute, Ljubljana}
\affiliation{Kanagawa University, Yokohama}
\affiliation{Korea University, Seoul}
\affiliation{Kyungpook National University, Taegu}
\affiliation{Swiss Federal Institute of Technology of Lausanne, EPFL, Lausanne}
\affiliation{University of Ljubljana, Ljubljana}
\affiliation{University of Maribor, Maribor}
\affiliation{University of Melbourne, Victoria}
\affiliation{Nagoya University, Nagoya}
\affiliation{Nara Women's University, Nara}
\affiliation{National Kaohsiung Normal University, Kaohsiung}
\affiliation{National United University, Miao Li}
\affiliation{Department of Physics, National Taiwan University, Taipei}
\affiliation{H. Niewodniczanski Institute of Nuclear Physics, Krakow}
\affiliation{Nihon Dental College, Niigata}
\affiliation{Niigata University, Niigata}
\affiliation{Osaka City University, Osaka}
\affiliation{Osaka University, Osaka}
\affiliation{Panjab University, Chandigarh}
\affiliation{Peking University, Beijing}
\affiliation{Princeton University, Princeton, New Jersey 08545}
\affiliation{University of Science and Technology of China, Hefei}
\affiliation{Sungkyunkwan University, Suwon}
\affiliation{University of Sydney, Sydney NSW}
\affiliation{Tata Institute of Fundamental Research, Bombay}
\affiliation{Toho University, Funabashi}
\affiliation{Tohoku Gakuin University, Tagajo}
\affiliation{Tohoku University, Sendai}
\affiliation{Department of Physics, University of Tokyo, Tokyo}
\affiliation{Tokyo Institute of Technology, Tokyo}
\affiliation{Tokyo Metropolitan University, Tokyo}
\affiliation{University of Tsukuba, Tsukuba}
\affiliation{Virginia Polytechnic Institute and State University, Blacksburg, Virginia 24061}
\affiliation{Yonsei University, Seoul}
  \author{K.~Abe}\affiliation{High Energy Accelerator Research Organization (KEK), Tsukuba} 
  \author{K.~Abe}\affiliation{Tohoku Gakuin University, Tagajo} 
  \author{H.~Aihara}\affiliation{Department of Physics, University of Tokyo, Tokyo} 
  \author{Y.~Asano}\affiliation{University of Tsukuba, Tsukuba} 
  \author{V.~Aulchenko}\affiliation{Budker Institute of Nuclear Physics, Novosibirsk} 
  \author{T.~Aushev}\affiliation{Institute for Theoretical and Experimental Physics, Moscow} 
  \author{S.~Bahinipati}\affiliation{University of Cincinnati, Cincinnati, Ohio 45221} 
  \author{A.~M.~Bakich}\affiliation{University of Sydney, Sydney NSW} 
  \author{Y.~Ban}\affiliation{Peking University, Beijing} 
  \author{I.~Bedny}\affiliation{Budker Institute of Nuclear Physics, Novosibirsk} 
% ...
\collaboration{The Belle Collaboration}